\documentclass{acm_sen_article}

\usepackage[lmargin=1.27cm,rmargin=1.27cm,bmargin=1.8cm,tmargin=1.9cm]{geometry}
\usepackage{url}
\usepackage{graphicx}
\usepackage{listings}
\usepackage{array}
\usepackage{multirow}
\usepackage{breakcites}
\usepackage{subfig}
\usepackage{morefloats}
\usepackage{caption}
\usepackage{xcolor}
\usepackage{doi}
\usepackage{hyperref}
\usepackage{paralist}

\begin{document}

\title{Benchmarking of Java Verification Tools\\ at the Software Verification Competition (SV-COMP)}


\numberofauthors{3}

\author{
\alignauthor
Lucas Cordeiro\\
       \affaddr{University of Manchester}\\
       \affaddr{Manchester, UK}\\
       \email{lucas.cordeiro@\\manchester.ac.uk}
\alignauthor
Daniel Kroening\\
       \affaddr{University of Oxford}\\
       \affaddr{Oxford, UK}\\
       \email{kroening@cs.ox.ac.uk}
\and
\alignauthor
Peter Schrammel\\
       \affaddr{University of Sussex}\\
       \affaddr{Brighton, UK}\\
       \email{p.schrammel@sussex.ac.uk}
}
\maketitle

\begin{abstract}
  Empirical evaluation of verification tools by benchmarking
  is a common method in software verification research.
  The Competition on Software Verification (SV-COMP) aims at
  standardization and reproducibility of benchmarking within the
  software verification community on an annual basis, through comparative
  evaluation of fully automatic software verifiers for C programs.
  Building upon this success, here we describe how
  to re-use the ecosystem developed around SV-COMP for benchmarking
  Java verification tools. We provide a detailed description of the
  rules for benchmark verification tasks, the integration of
  new tools into SV-COMP's benchmarking framework and
  also give experimental results of a benchmarking run on state-of-the-art
  Java verification tools, JPF, SPF, JayHorn and JBMC.
\end{abstract}

\section{Introduction}

The complexity of software in smartphones and enterprise applications has
dramatically increased over the last years.  In particular, mobile
applications based on the Android OS have gained popularity in the consumer
electronics industry, reaching nearly 87\% market share~\cite{idc2018}; the
Android OS essentially consists of a large set of libraries (approximately
13 million lines of code), containing both Java code and native code, which
thus require methods to verify its security properties (e.g., sensitive data
leakage)~\cite{BaiYWBSLDV18}.  Similarly, Java remains popular in business
applications (server-side), mainly owing to the existence of several robust
frameworks (e.g., Spring~\cite{spring2018}); therefore, verification of Java
enterprise applications is also of particular interest.

Technology companies such as Facebook and Amazon increasingly invest effort
and time to develop efficient and effective verification methods as testing
alternatives~\cite{OHearn18, CookKKTTT18}, to check correctness of some
aspects of their systems with the goal to improve robustness and
security~\cite{Durumeric:2014:MH:2663716.2663755}.  Although there are
several software verification tools for Java programs (e.g.,
Bandera~\cite{IosifDH05},
JPF~\cite{DBLP:conf/kbse/VisserHBP00},
SPF~\cite{DBLP:conf/tacas/AnandPV07},
JayHorn~\cite{DBLP:conf/cav/KahsaiRSS16} and
JBMC~\cite{DBLP:conf/cav/CordeiroKKST18}), they are typically very difficult
to compare in practice, mainly due to the lack of (1)~a common set of
benchmarks and (2)~methods to standardize and reproduce the empirical
evaluations.

\paragraph{The Software Verification Competition} SV-COMP is one of the main
initiatives targeted at the evaluation of new software verification methods,
technologies, and tools~\cite{DBLP:conf/tacas/Beyer17}.  It has been running
since $2012$ as part of the International Conference on Tools and Algorithms
for the Construction and Analysis of Systems (TACAS).  Its main focus has
been on evaluating different verification (and testing) tools for C~programs. 
Currently, there are some powerful Java verifiers available, but there is no
standard procedure to compare them fairly.  One of the main difficulties to
conduct such a comparison is the lack of a standard set of Java benchmarks
and respective benchmarking infrastructure to obtain reliable, reproducible
and accurate results.

The main contribution of this paper is to define a standard Java benchmark
format and respective benchmarking infrastructure, which can drive the
verification community to effectively evaluate state-of-the-art software
verification tools for Java programs with the goal to achieve comparability
and reproducibility.  In particular, we collect and harmonize an initial
set of Java benchmarks from different
sources~\cite{DBLP:conf/cav/KahsaiRSS16, DBLP:conf/cav/CordeiroKKST18,
DBLP:conf/tacas/AnandPV07, DBLP:conf/kbse/PasareanuR10} and re-use existing
benchmarking infrastructure~\cite{Beyer2017}, so that we allow the community
to get beyond research prototypes to usable
tools~\cite{DBLP:conf/atva/AlglaveDKT11}.  This can lead to further progress
in the area of verification of Java programs and raise interest in applying these tools to industrial systems.

\paragraph{Java verification tools}
Here, we consider the following tools:

JBMC~\cite{DBLP:conf/cav/CordeiroKKST18}\footnote{Available at
  \url{https://www.cprover.org/jbmc/}} is based on the C
Bounded Model Checker (CBMC)~\cite{DBLP:conf/tacas/ClarkeKL04}
to verify Java bytecode.  JBMC consists of a frontend for parsing Java
bytecode and a Java operational model (JOM), which is an exact but
verification-friendly model of the standard Java libraries.
A distinct feature of JBMC is the
use of Bounded Model Checking (BMC)~\cite{handbook09} in combination with
Boolean Satisfiability and Satisfiability Modulo Theories
(SMT)~\cite{BarrettSST09} and full symbolic state-space exploration, which
allows JBMC to perform a bit-accurate verification of Java programs.

\emph{Symbolic PathFinder} (SPF)\footnote{Available at
  \url{https://github.com/symbolicpathfinder}} is a
symbolic~\cite{DBLP:conf/tacas/AnandPV07,DBLP:conf/kbse/PasareanuR10}
software model checking extension of \emph{Java PathFinder}
(JPF)\footnote{Available at \url{https://github.com/javapathfinder}},
an explicit-state model checker for Java
bytecode~\cite{DBLP:conf/kbse/VisserHBP00}.  JPF is used to find and
explain defects, collect runtime information as coverage metrics,
deduce test vectors, and create corresponding test drivers for Java
programs.  JPF checks for property violations such as deadlocks or
unhandled exceptions along all potential execution paths as well as
user-specified assertions.

\emph{JayHorn}\footnote{Available at
  \url{https://github.com/jayhorn/jayhorn}} is a verifier for Java
bytecode~\cite{DBLP:conf/cav/KahsaiRSS16} that uses the Java
optimization framework Soot~\cite{Vallee-Rai:1999:SJB:781995.782008}
as a front-end and then produces a set of constrained Horn clauses to
encode the verification condition (VC). JayHorn is able to check for
user-specified assertions and is sound for Java programs that use a
single thread, have no dynamic class loading and complex static
initializers.

\paragraph{Overview} This paper proposes a Java Category for the Software
Verification Competition (SV-COMP).  First, we describe in detail the
definition and set up of the category.  Then, we report on the integration
of the tools mentioned above into the SV-COMP benchmarking
infrastructure.  Finally, we give the experimental results that we obtained
by benchmarking the tools on the benchmarks that we collected.

\section{A Java Category in SV-COMP}

We describe the proposed Java category in SV-COMP.  In
particular, we define the structure and meaning of
verification tasks, the properties to be verified, the execution
environment and how verification results could be evaluated.%
\footnote{We describe the Java category as executed in SV-COMP 2019.
The original proposal can be found in \cite{CKS18}.}

\subsection{Definition of Verification Task}\label{sec:vtask}

A verification task consists of a Java program and a specification. A
verification run is a non-interactive execution of a competition
candidate, i.e., a verifier, on a single verification task, in order
to check whether the program satisfies its specification.
According to the current SV-COMP rules,\footnote{A detailed description of the current rules can be found in
\url{https://sv-comp.sosy-lab.org/2019/rules.php}} the result of a verification run is a triple 
(\textit{answer}, \textit{witness}, \textit{time}). \textit{Answer} is one of the following 
outcomes given in Table~\ref{tab:outcomes}.
\begin{table}[h]
\begin{center}
\caption{Definition of a Verification Result in SV-COMP~\cite{DBLP:conf/tacas/Beyer17}.}
\label{tab:outcomes}
\begin{tabular}{|l|p{0.34\textwidth}|}
\hline
TRUE & The specification is satisfied (i.e., there is no path that violates
the specification). \\
\hline
FALSE & The specification is violated (i.e., there exists a path that violates
the specification). \\
\hline
UNKNOWN & The tool cannot decide the problem or terminates by a tool crash,
time-out, or out-of-memory (i.e., the competition candidate does not
succeed in computing an answer TRUE or FALSE). \\
\hline
\end{tabular}
\end{center}
\end{table}

\textit{Time} is the consumed CPU time until the verifier terminates. It
includes the consumed CPU time of all processes that the verifier
starts. If \textit{time} is equal to or larger than the time limit, then the
verifier is terminated and the \textit{answer} is set to ``timeout'' (and
interpreted as UNKNOWN).

Witness checking as previously described
in~\cite{DBLP:conf/sigsoft/0001DDH16}) represents an important feature
to validate verification results given by verifiers.
At the moment there is no witness checking in the Java category, though.

The Java verification tasks are partitioned into categories, which are defined
in category-set files. At the moment there is only one category in
Java, \textit{ReachSafety}, which is concerned with specifications
that consider an \textsf{assert(\textit{condition})} statement in the
verification task whose non-violation must be proven or
refuted.  

\begin{figure*}
  \centering\includegraphics[width=\textwidth]{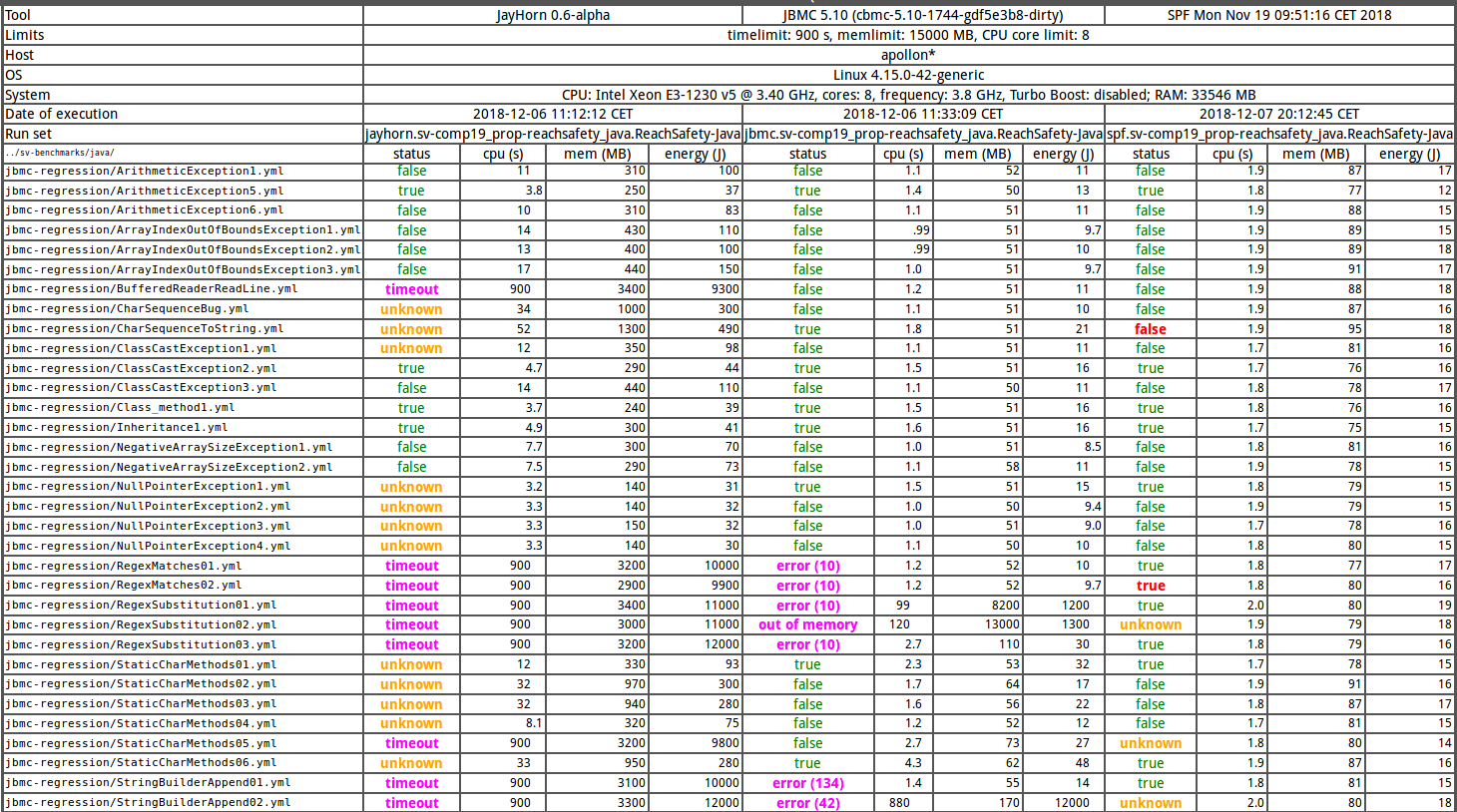}
  \caption{\label{fig:results}
    Per benchmark comparison table as produced by BenchExec~\cite{Beyer2017}. The detailed description of the scores of each tool can be found in \url{https://sv-comp.sosy-lab.org/2019/results/results-verified/META_JavaOverall.table.html}.}
\end{figure*}

\subsection{Benchmark Verification Tasks}

All Java verification tasks used in the competition must be part of
the SV-COMP benchmark
collection\footnote{\url{https://github.com/sosy-lab/sv-benchmarks}}
prior to the benchmark contribution deadline (typically in September).
The competition candidates can ``train'', i.e., run and tune, their
verifiers on the verification tasks until the tool submission deadline
(typically in November). Misclassified benchmarks can be corrected
during this training period. Contentious benchmarks 
are excluded from the competition once they are identified. 
SV-COMP does not use verification tasks without
training; this is particularly important for Java since it has many features 
of which verifiers only support a subset; participants should know before 
the competition which features they are expected to support.

\paragraph{Benchmark structure}

Verification tasks are grouped in directories depending on their
source, e.g., \textsf{jbmc-regression}. Within these directories, each
verification task consists of a YAML file,
e.g. \textsf{StringValueOf01} in the format defined by BenchExec for
task-definition files. These YAML files define the list of input files
(Java sources) of a task and the expected verdict for each possible
property.

The programs are assumed to be written in Java 1.8. Verification tools
that require the sources to be compiled can use any Java 1.8 compiler.
The verification task need to be compilable by putting all
\textsf{.java} files in directories listed as input files in the YAML
file on the sourcepath of a Java 8 compiler.

The program may call the Java standard library (\textsf{java.*},
\textsf{javax.*}). The sources of other dependencies must be added to
the source tree together with their respective licenses.  In order to
allow tools that analyze Java source code to participate, we do not
permit \textsf{.jar} files as dependencies (except the Java standard
library).

BenchExec will pass the paths that are listed as input files in a
task-definition file to the tool-info module, which can pass them to
the verifier or for example expand them to a list of single
\textsf{.java} files, depending on what the verifier needs. If a
verification tool requires \textsf{.class} files or a \textsf{.jar}
file as input it should use regular Java utilities to create these
artifacts (in a wrapper script if necessary).

The benchmark must have a \textsf{Main} class with a \textsf{public
  static void main(String[])} method in the root package. The
\textsf{Main.java} file must have a copyright header indicating the
source of the benchmark and its license.

Potential competition participants are invited to submit verification
tasks until the benchmark contribution deadline by submitting a Pull
Request to the benchmark collection repository.\footnote{\url{https://github.com/sosy-lab/sv-benchmarks}}
Verification tasks must comply with the aforementioned format.
New proposed categories will be included if at least three different
tools or teams participate in the category (i.e., not the same tool
twice with a different configuration).
In the following, we list a few conventions that are used in
the Java verification tasks.

\paragraph{Assertions}
For checking (un)reachability, we use the \textsf{assert} keyword provided
in the Java language. It is assumed that the \textsf{AssertionError} thrown
on violation of the assertion always leads to abortion of the
program, i.e., it is not caught in the program.
In future, further properties could be defined by different types of
uncaught errors/exceptions.

\paragraph{Nondeterminism}
The only admissible source of nondeterminism are the return values of
the methods defined in the \textsf{org.sosy\_lab.}
\textsf{sv\_benchmarks.Verifier} class, provided in
\textsf{java/common/org/sosy\_lab}
\textsf{/sv\_benchmarks/Verifier.java} in the \textsf{sv-benchmarks}
repository. In order to make the benchmarks compilable,
\textsf{../common/} needs to be added to the \textsf{input\_files}
property of the benchmark's YAML file.

The methods in \textsf{org.sosy\_lab.sv\_benchmarks.Verifier} call
methods of the \textsf{java.util.Random class}. The rationale is to provide
straightforward compatibility with verifiers that implement a
nondeterministic semantics for the \textsf{java.util.Random class}, i.e. the
methods in \textsf{java.util.Random} are expected to return a nondeterministic
value instead of a random value, but satisfying the same constraints
on their value range.

\textsf{org.sosy\_lab.sv\_benchmarks.Verifier} also provides an assume
method, which is defined as \textsf{Runtime.getRuntime().halt(1)}. It
is recommended to use assume or return (if in the entry point method)
to restrict the nondeterminism as they do not impact the termination
behavior of a program. For example, using \textsf{while(!condition);} would
make any program with such assumptions be classified non-terminating
when a potential \textit{Java Termination} category might be introduced in
future.

\paragraph{Operating System Model}
Any library methods that make system calls are not allowed in
verification tasks.
Exceptions with well-defined behaviors can be explicitly granted if
they allow a wider range of benchmarks to be included in the
collection. For instance, \textsf{new java.util.Date()} could be
defined to create a \textsf{Date} object with a nondeterministic
timestamp.

\subsection{Properties}

In SV-COMP, the specification to be verified for a program is given
in \textsf{.prp} files.
The definitions in these \textsf{.prp} have been designed
for extensibility in order to allow new properties for future
categories to be specified. For instance, in the C categories, four
different properties are in use.

For our Java category, the definition in
\textsf{java/properties/assert.prp} states \textsf{CHECK(
  init(Main.main()), LTL(G assert) )}.  Here,
\textsf{init(Main.} \textsf{main())} gives the initial states of the program by
a call of the \textsf{public static void main(String[])} function of
the \textsf{Main} class.  \textsf{LTL(f)} specifies that formula
\textsf{f} holds at every initial state of the program. In particular,
the linear-time temporal logic (LTL) operator \textsf{G f} means that
\textsf{f} globally holds.  The proposition \textsf{assert} is true if
all \textsf{assert} statements in the program hold.
\pagebreak

\subsection{Competition Environment and Requirements}

In the SV-COMP environment, each software verifier is assumed to run
on a machine with a GNU/Linux operating system (x86\_64-linux, Ubuntu
18.04). SV-COMP also sets three resource limits to evaluate each
software verifier, which are: $(i)$ a memory limit of $15$ GB of RAM,
$(ii)$ a runtime limit of $900$ seconds of CPU time, and $(iii)$ a
limit to $8$ processing units of a CPU.
Note that if a software verifier does not consume CPU time, then it is
killed after $900$ seconds of wall clock time, and the resulting
runtime is set to $900$ seconds.  For the Java category, OpenJDK 1.8
is assumed to be installed on the competition machines.
The modest resource requirements have been chosen in order to allow
everybody to reproduce the competition results on a reasonably sized
machine.

\begin{figure}
  \centering\includegraphics[width=0.4\textwidth]{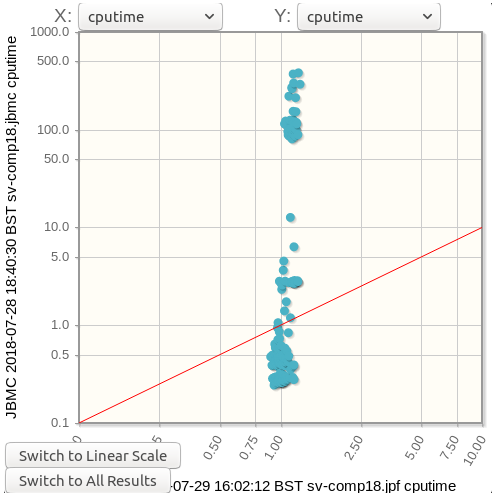}\\[0.8ex]
  \caption{\label{fig:scatter}
    Example of a scatter plot comparing SPE and JBMC as produced by BenchExec~\cite{DBLP:conf/spin/0001LW15a}.}
\end{figure}

\subsection{Evaluation by Scores and Runtime}

SV-COMP has strict rules to evaluate the verification results provided
by each software verifier.  In particular, each verifier is heavily
penalized if they produce an incorrect result for a specific verification
task with the goal to favor correctness.
The scores are assigned to each software verifier according to Table~\ref{tab:scores}.\footnote{\url{https://sv-comp.sosy-lab.org/2019/rules.php}}

\begin{table}[h]
\begin{center}
\caption{Evaluation by Scores and Runtime in SV-COMP~\cite{DBLP:conf/tacas/Beyer17}.}
\label{tab:scores}
\begin{tabular}{|l|p{0.09\textwidth}|p{0.27\textwidth}|}
\hline
Points &	Answer & Description \\
\hline
0 & UNKNOWN & Failure to compute verification result, out of resources, program crash. \\
\hline
+1 &	FALSE\quad correct &	The error in the program was found. \\
\hline
-16 &	FALSE\quad incorrect &	An error is reported for a program that fulfills the specification (false alarm, incomplete analysis). \\
\hline
+2 &	TRUE\quad correct & 	The program was analyzed to be free of errors. \\
\hline
-32 &	TRUE\quad incorrect &	The program had an error but the competition candidate did not find it (missed bug, unsound analysis). \\
\hline
\end{tabular}
\end{center}
\end{table}

The higher score and penalty for the TRUE case is justified because it
is usually more difficult to prove a program correct than to find a
bug~\cite{DBLP:conf/tacas/Beyer17}.

\section{Integration into the Competition Infrastructure}\label{sec:bexec}

BenchExec\footnote{\url{https://github.com/sosy-lab/benchexec}}~\cite{Beyer2017}
is the framework used in SV-COMP for reliable benchmarking and
resource measurement. It can be easily installed to run experimental
comparisons and to reproduce competition results.
For the Java category, we extended the framework by introducing a new
\textsf{assert} proposition for specifying properties.

Integrating a new tool into the framework requires the addition of two files:
\begin{compactitem}
\item The \textit{tool-info module} is a Python module located in the \textsf{benchexec/tools} directory that implements the tool interface to connect a verifier to BenchExec. Essentially, it must provide functions for running the verifier with a given verification task and to translate the tool output into an \textit{answer} TRUE, FALSE or UNKNOWN (see Section~\ref{sec:vtask}).
\item The \textit{benchmark definition}\footnote{\url{https://github.com/sosy-lab/sv-comp}} is an XML file that specifies which categories can be run with a given verifier and which tool command line options to use.
\end{compactitem}

Here, we have implemented and added these files for the tools 
under consideration (i.e., JPF, SPF, JBMC and JayHorn).
After installing a verifier (e.g., SPF) in the base directory of
BenchExec, it can be run with the command \textsf{bin/benchexec
  spf.xml}.  There are various options to run subsets of the
verification tasks and overriding time and memory limits, for
instance.

\section{Java Benchmark Collection}

Previously, there existed $64$ ``minepump'' benchmarks in the SV-COMP
repository from earlier attempts to run a Java category; these benchmarks
were already classified as ``safe'' and ``unsafe'' by the community.  
Beyond these few files, there was no standard benchmark suite for Java verification
available in the community.\footnote{There is a community effort in collecting Java 
benchmarks in \url{http://sir.unl.edu}, but they are not currently classified.}
Therefore, we took the entire JBMC regression test suite (``jbmc-regression''), consisting of $177$
benchmarks (including known bugs and hard benchmarks that JBMC cannot
yet handle); these benchmarks test common Java features (e.g.,
polymorphism, exceptions, arrays, and strings) and they were classified
by the JBMC developers.  We also used $23$
benchmarks (``jayhorn-recursive'') taken from the JayHorn
repository~\cite{DBLP:conf/cav/KahsaiRSS16}. These are mainly C
benchmarks from the \textsf{recursive} category that have been
translated into Java by keeping the original classification from SV-COMP.  
Additionally, we have extracted $104$ benchmarks from the JPF regression test
suite~\cite{DBLP:conf/tacas/AnandPV07} (``jpf-regression''); for these particular 
benchmarks, we have manually inspected and classified them as ``safe'' and ``unsafe''.
Table~\ref{tab:bench} summarizes the characteristics of the benchmark sets.%

\begin{table}[h]
\begin{center}
\caption{Characteristics of the Java Benchmark Sets.}
\label{tab:bench}
\begin{tabular}{lrrrr}
\hline
benchmark set & total & safe & unsafe & avg.~LOC \\
\hline
jbmc-regression      & 177 & 89 & 88 & 25 \\
jpf-regression       & 104 & 52 & 52 & 52 \\
jayhorn-recursive &  23 & 14 &  9 & 35 \\
minepump  &  64 &  8 & 56 & 62 \\
\hline
total     & 368 & 163 & 205 & 40 \\
\hline
\end{tabular}
\end{center}
\end{table}

These benchmarks are a good start to launch a Java
category, but are not yet fully representative for the breadth of
challenges that we face in verifying Java programs. We will rely on
the community to contribute and continuously enrich the collection of
Java benchmarks in future editions of the competition.

\section{Benchmarking Results}

The results of running a verifier using BenchExec as explained in
Section~\ref{sec:bexec} are collected in a timestamped format in the
\textsf{results} directory with the BenchExec base directory. This
contains a \textsf{.zip} file with the log files and a
\textsf{.xml.bz2} file with the results in a structured format.  One
or more of the latter files (potentially of different tools) can be
passed to \textsf{bin/table-generator} in order to generate an HTML
report that compares the benchmarking runs.

\begin{figure*}
  \centering\includegraphics[width=0.7\textwidth]{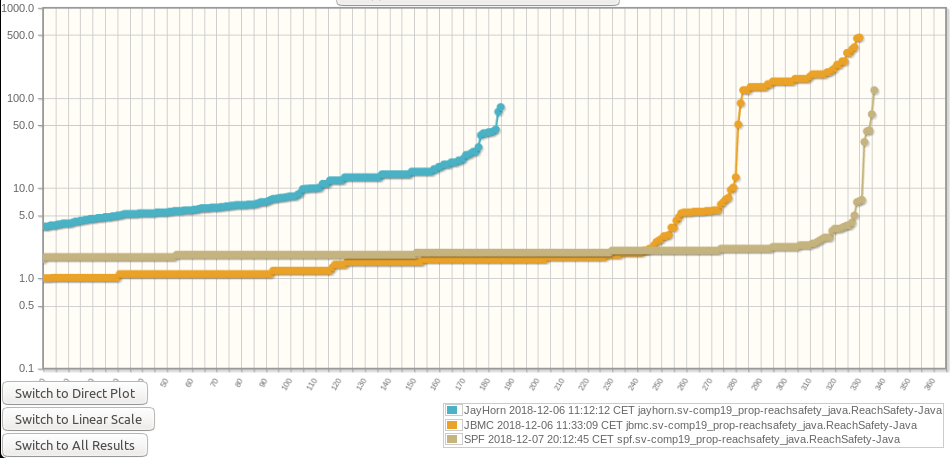}
  \caption{\label{fig:quantile}
    Quantile plot as produced by BenchExec~\cite{Beyer2017}.}
\end{figure*}

A part of this report is shown in Figure~\ref{fig:results}.\footnote{The full results are available at
\url{https://sv-comp.sosy-lab.org/2019/results/results-verified/META_JavaOverall.table.html}}.
%
%
The HTML report allows to filter rows and columns and display the most
important comparison charts, such as scatter plots ---
Figure~\ref{fig:scatter} compares the CPU time of SPF and JBMC, for
example --- and quantile (``cactus'') plots as depicted in
Figure~\ref{fig:quantile}.
The latter plot shows the cumulative time (y-axis) required for a
verifier to solve its $n$ fastest benchmarks (x-axis). This allows us
to compare the scaling behavior of the tools, i.e., the longer a
graph extends to the right the more verification tasks were solved by
the tool, the closer to the bottom the faster it is.

\section{Conclusions}

We described our proposal to run the first Java category in SV-COMP 2019,
given that it is currently focused on evaluating C software verifiers
only.  In particular, we defined the structure and meaning of
verification tasks, the properties to be verified, the execution
environment and how verifiers are integrated into the benchmarking
framework and how verification results, produced by each Java verifier
are evaluated.
SV-COMP is one of the most successful software verification
competitions, which is annually held by TACAS. 
Although the first edition of SV-COMP took place in 2012 and
has been a successful so far, there has been no verification track to
evaluate software verifiers targeted for Java programs.
As a next step, Java verifiers need to be extended in order to produce
witness files (for violation and correctness) that adhere to the
witness exchange format defined by
SV-COMP~\cite{DBLP:conf/sigsoft/0001DDH16}.  In this respect, witness
checkers for Java verifiers also need to developed in order to check
validity of the verification results provided by each
verifier~\cite{DBLP:conf/tap/0001DLT18}.




\newcommand{\BMCxmlcomment}[1]{}

\BMCxmlcomment{

<refgrp>

<bibl id="B1">
  <title><p>Smartphone {OS} market share, 2016 Q3</p></title>
  <aug>
    <au><cnm>IDC</cnm></au>
  </aug>
  <pubdate>2018</pubdate>
  <url>http://www.idc.com/prodserv/smartphone-os-market-share.jsp</url>
</bibl>

<bibl id="B2">
  <title><p>Towards Model Checking {Android} Applications</p></title>
  <aug>
    <au><snm>Bai</snm><fnm>G</fnm></au>
    <au><snm>Ye</snm><fnm>Q</fnm></au>
    <au><snm>Wu</snm><fnm>Y</fnm></au>
    <au><snm>Botha</snm><fnm>H</fnm></au>
    <au><snm>Sun</snm><fnm>J</fnm></au>
    <au><snm>Liu</snm><fnm>Y</fnm></au>
    <au><snm>Dong</snm><fnm>JS</fnm></au>
    <au><snm>Visser</snm><fnm>W</fnm></au>
  </aug>
  <source>{IEEE} Trans. Software Eng.</source>
  <pubdate>2018</pubdate>
  <volume>44</volume>
  <issue>6</issue>
  <fpage>595</fpage>
  <lpage>-612</lpage>
</bibl>

<bibl id="B3">
  <title><p>Why is Spring more popular than other {Java}
  frameworks?</p></title>
  <aug>
    <au><snm>Gupta</snm><fnm>K</fnm></au>
  </aug>
  <pubdate>2018</pubdate>
  <url>https://www.freelancinggig.com/blog/2018/04/26/spring-popular-java-frameworks/</url>
</bibl>

<bibl id="B4">
  <title><p>Continuous Reasoning: Scaling the impact of formal
  methods</p></title>
  <aug>
    <au><snm>O'Hearn</snm><fnm>PW</fnm></au>
  </aug>
  <source>{LICS}</source>
  <pubdate>2018</pubdate>
  <fpage>13</fpage>
  <lpage>-25</lpage>
</bibl>

<bibl id="B5">
  <title><p>Model Checking Boot Code from {AWS} Data Centers</p></title>
  <aug>
    <au><snm>Cook</snm><fnm>B</fnm></au>
    <au><snm>Khazem</snm><fnm>K</fnm></au>
    <au><snm>Kroening</snm><fnm>D</fnm></au>
    <au><snm>Tasiran</snm><fnm>S</fnm></au>
    <au><snm>Tautschnig</snm><fnm>M</fnm></au>
    <au><snm>Tuttle</snm><fnm>MR</fnm></au>
  </aug>
  <source>{CAV}</source>
  <series><title><p>LNCS</p></title></series>
  <pubdate>2018</pubdate>
  <volume>10982</volume>
  <fpage>467</fpage>
  <lpage>-486</lpage>
</bibl>

<bibl id="B6">
  <title><p>The Matter of {Heartbleed}</p></title>
  <aug>
    <au><snm>Durumeric</snm><fnm>Z</fnm></au>
    <au><snm>Kasten</snm><fnm>J</fnm></au>
    <au><snm>Adrian</snm><fnm>D</fnm></au>
    <au><snm>Halderman</snm><fnm>JA</fnm></au>
    <au><snm>Bailey</snm><fnm>M</fnm></au>
    <au><snm>Li</snm><fnm>F</fnm></au>
    <au><snm>Weaver</snm><fnm>N</fnm></au>
    <au><snm>Amann</snm><fnm>J</fnm></au>
    <au><snm>Beekman</snm><fnm>J</fnm></au>
    <au><snm>Payer</snm><fnm>M</fnm></au>
    <au><snm>Paxson</snm><fnm>V</fnm></au>
  </aug>
  <source>IMC</source>
  <pubdate>2014</pubdate>
  <fpage>475</fpage>
  <lpage>-488</lpage>
</bibl>

<bibl id="B7">
  <title><p>Translating Java for Multiple Model Checkers: The Bandera
  Back-End</p></title>
  <aug>
    <au><snm>Iosif</snm><fnm>R</fnm></au>
    <au><snm>Dwyer</snm><fnm>MB</fnm></au>
    <au><snm>Hatcliff</snm><fnm>J</fnm></au>
  </aug>
  <source>Formal Methods in System Design</source>
  <pubdate>2005</pubdate>
  <volume>26</volume>
  <issue>2</issue>
  <fpage>137</fpage>
  <lpage>-180</lpage>
</bibl>

<bibl id="B8">
  <title><p>Model Checking Programs</p></title>
  <aug>
    <au><snm>Visser</snm><fnm>W</fnm></au>
    <au><snm>Havelund</snm><fnm>K</fnm></au>
    <au><snm>Brat</snm><fnm>GP</fnm></au>
    <au><snm>Park</snm><fnm>S</fnm></au>
  </aug>
  <source>{ASE}</source>
  <pubdate>2000</pubdate>
  <fpage>3</fpage>
  <lpage>-12</lpage>
</bibl>

<bibl id="B9">
  <title><p>{JPF-SE:} {A} Symbolic Execution Extension to {Java}
  {PathFinder}</p></title>
  <aug>
    <au><snm>Anand</snm><fnm>S</fnm></au>
    <au><snm>Pasareanu</snm><fnm>CS</fnm></au>
    <au><snm>Visser</snm><fnm>W</fnm></au>
  </aug>
  <source>{TACAS}</source>
  <series><title><p>LNCS</p></title></series>
  <pubdate>2007</pubdate>
  <volume>4424</volume>
  <fpage>134</fpage>
  <lpage>-138</lpage>
</bibl>

<bibl id="B10">
  <title><p>{JayHorn}: {A} Framework for Verifying {Java} programs</p></title>
  <aug>
    <au><snm>Kahsai</snm><fnm>T</fnm></au>
    <au><snm>R{\"{u}}mmer</snm><fnm>P</fnm></au>
    <au><snm>Sanchez</snm><fnm>H</fnm></au>
    <au><snm>Sch{\"{a}}f</snm><fnm>M</fnm></au>
  </aug>
  <source>{CAV}</source>
  <series><title><p>LNCS</p></title></series>
  <pubdate>2016</pubdate>
  <volume>9779</volume>
  <fpage>352</fpage>
  <lpage>-358</lpage>
</bibl>

<bibl id="B11">
  <title><p>{JBMC:} {A} Bounded Model Checking Tool for Verifying {Java}
  Bytecode</p></title>
  <aug>
    <au><snm>Cordeiro</snm><fnm>LC</fnm></au>
    <au><snm>Kesseli</snm><fnm>P</fnm></au>
    <au><snm>Kroening</snm><fnm>D</fnm></au>
    <au><snm>Schrammel</snm><fnm>P</fnm></au>
    <au><snm>Trt{\'{\i}}k</snm><fnm>M</fnm></au>
  </aug>
  <source>{CAV}</source>
  <series><title><p>LNCS</p></title></series>
  <pubdate>2018</pubdate>
  <volume>10981</volume>
  <fpage>183</fpage>
  <lpage>-190</lpage>
</bibl>

<bibl id="B12">
  <title><p>Software Verification with Validation of Results (Report on
  {SV-COMP} 2017)</p></title>
  <aug>
    <au><snm>Beyer</snm><fnm>D</fnm></au>
  </aug>
  <source>{TACAS}</source>
  <series><title><p>LNCS</p></title></series>
  <pubdate>2017</pubdate>
  <volume>10206</volume>
  <fpage>331</fpage>
  <lpage>-349</lpage>
</bibl>

<bibl id="B13">
  <title><p>Symbolic {PathFinder}: symbolic execution of {Java}
  bytecode</p></title>
  <aug>
    <au><snm>Pasareanu</snm><fnm>CS</fnm></au>
    <au><snm>Rungta</snm><fnm>N</fnm></au>
  </aug>
  <source>{ASE}</source>
  <pubdate>2010</pubdate>
  <fpage>179</fpage>
  <lpage>-180</lpage>
</bibl>

<bibl id="B14">
  <title><p>Reliable benchmarking: requirements and solutions</p></title>
  <aug>
    <au><snm>Beyer</snm><fnm>D</fnm></au>
    <au><snm>L{\"o}we</snm><fnm>S</fnm></au>
    <au><snm>Wendler</snm><fnm>P</fnm></au>
  </aug>
  <source>International Journal on Software Tools for Technology Transfer (to
  appear)</source>
  <pubdate>2017</pubdate>
</bibl>

<bibl id="B15">
  <title><p>Making Software Verification Tools Really Work</p></title>
  <aug>
    <au><snm>Alglave</snm><fnm>J</fnm></au>
    <au><snm>Donaldson</snm><fnm>AF</fnm></au>
    <au><snm>Kroening</snm><fnm>D</fnm></au>
    <au><snm>Tautschnig</snm><fnm>M</fnm></au>
  </aug>
  <source>{ATVA}</source>
  <series><title><p>LNCS</p></title></series>
  <pubdate>2011</pubdate>
  <volume>6996</volume>
  <fpage>28</fpage>
  <lpage>-42</lpage>
</bibl>

<bibl id="B16">
  <title><p>A Tool for Checking {ANSI-C} Programs</p></title>
  <aug>
    <au><snm>Clarke</snm><fnm>EM</fnm></au>
    <au><snm>Kroening</snm><fnm>D</fnm></au>
    <au><snm>Lerda</snm><fnm>F</fnm></au>
  </aug>
  <source>{TACAS}</source>
  <series><title><p>LNCS</p></title></series>
  <pubdate>2004</pubdate>
  <volume>2988</volume>
  <fpage>168</fpage>
  <lpage>-176</lpage>
</bibl>

<bibl id="B17">
  <title><p>Bounded Model Checking</p></title>
  <aug>
    <au><snm>Biere</snm><fnm>A</fnm></au>
  </aug>
  <source>Handbook of Satisfiability</source>
  <series><title><p>Frontiers in Artificial Intelligence and
  Applications</p></title></series>
  <pubdate>2009</pubdate>
  <volume>185</volume>
  <fpage>457</fpage>
  <lpage>-481</lpage>
</bibl>

<bibl id="B18">
  <title><p>Satisfiability Modulo Theories</p></title>
  <aug>
    <au><snm>Barrett</snm><fnm>C</fnm></au>
    <au><snm>Sebastiani</snm><fnm>R</fnm></au>
    <au><snm>Seshia</snm><fnm>SA</fnm></au>
    <au><snm>Tinelli</snm><fnm>C</fnm></au>
  </aug>
  <source>Handbook of Satisfiability</source>
  <series><title><p>Frontiers in Artificial Intelligence and
  Applications</p></title></series>
  <pubdate>2009</pubdate>
  <volume>185</volume>
  <fpage>825</fpage>
  <lpage>-885</lpage>
</bibl>

<bibl id="B19">
  <title><p>Soot -- a {Java} Bytecode Optimization Framework</p></title>
  <aug>
    <au><snm>Vall{\'e}e Rai</snm><fnm>R</fnm></au>
    <au><snm>Co</snm><fnm>P</fnm></au>
    <au><snm>Gagnon</snm><fnm>E</fnm></au>
    <au><snm>Hendren</snm><fnm>L</fnm></au>
    <au><snm>Lam</snm><fnm>P</fnm></au>
    <au><snm>Sundaresan</snm><fnm>V</fnm></au>
  </aug>
  <source>CASCON</source>
  <pubdate>1999</pubdate>
  <fpage>13</fpage>
</bibl>

<bibl id="B20">
  <title><p>Benchmarking of {Java} Verification Tools at the Software
  Verification Competition {(SV-COMP)}</p></title>
  <aug>
    <au><snm>Cordeiro</snm><fnm>LC</fnm></au>
    <au><snm>Kroening</snm><fnm>D</fnm></au>
    <au><snm>Schrammel</snm><fnm>P</fnm></au>
  </aug>
  <source>{ACM} {SIGSOFT} Software Engineering Notes</source>
  <pubdate>2018</pubdate>
  <volume>43</volume>
  <issue>4</issue>
  <fpage>56</fpage>
</bibl>

<bibl id="B21">
  <title><p>Correctness witnesses: exchanging verification results between
  verifiers</p></title>
  <aug>
    <au><snm>Beyer</snm><fnm>D</fnm></au>
    <au><snm>Dangl</snm><fnm>M</fnm></au>
    <au><snm>Dietsch</snm><fnm>D</fnm></au>
    <au><snm>Heizmann</snm><fnm>M</fnm></au>
  </aug>
  <source>{FSE}</source>
  <pubdate>2016</pubdate>
  <fpage>326</fpage>
  <lpage>-337</lpage>
</bibl>

<bibl id="B22">
  <title><p>Benchmarking and Resource Measurement</p></title>
  <aug>
    <au><snm>Beyer</snm><fnm>D</fnm></au>
    <au><snm>L{\"{o}}we</snm><fnm>S</fnm></au>
    <au><snm>Wendler</snm><fnm>P</fnm></au>
  </aug>
  <source>{SPIN}</source>
  <series><title><p>LNCS</p></title></series>
  <pubdate>2015</pubdate>
  <volume>9232</volume>
  <fpage>160</fpage>
  <lpage>-178</lpage>
</bibl>

<bibl id="B23">
  <title><p>Tests from Witnesses -- Execution-Based Validation of Verification
  Results</p></title>
  <aug>
    <au><snm>Beyer</snm><fnm>D</fnm></au>
    <au><snm>Dangl</snm><fnm>M</fnm></au>
    <au><snm>Lemberger</snm><fnm>T</fnm></au>
    <au><snm>Tautschnig</snm><fnm>M</fnm></au>
  </aug>
  <source>{TAP}</source>
  <series><title><p>LNCS</p></title></series>
  <pubdate>2018</pubdate>
  <volume>10889</volume>
  <fpage>3</fpage>
  <lpage>-23</lpage>
</bibl>

</refgrp>
} 

\balancecolumns

\end{document}